\documentclass[letterpaper,10pt,natbib,cite]{article}
\usepackage{osameet2}

\usepackage{amsmath,amssymb}

\begin{document}

\title{Microrheology with Optical Tweezers of gel-like materials `\textit{is not an option}'!}

\author{Manlio Tassieri}
\address{Division of Biomedical Engineering, School of Engineering, University of Glasgow, Glasgow G12 8LT, UK}
\email{Manlio.Tassieri@glasgow.ac.uk}

\begin{abstract}
Optical tweezers have been successfully adopted as exceptionally sensitive transducers for microrheology studies of complex `\textit{fluids}'. Despite the general trend, a similar approach \emph{cannot} be adopted for microrheology studies of `\textit{gel-like}' materials, e.g. living cells.
\end{abstract}


\section{Introduction}
Since their advent in the $^{\prime}70s$~\cite{Ashkin:1970}, Optical Tweezers (OT) have revolutionised the field of micro-sensing\cite{Bowman:2013}.
Indeed, thanks to the property that a highly focused laser beam has to trap (in three dimensions) micron-sized dielectric objects suspended in a fluid, OT have been adopted as exceptionally sensitive transducers (i.e., able to resolve \textit{pN} forces and \textit{nm} displacements, with high temporal resolution, down to $\mu$\textit{sec}) to study a myriad of biological processes, e.g.~\cite{ISI:A1989U008400059, ISI:A1994NA03000055, ISI:000259700300007, ISI:A1987G470500040, ISI:A1997WZ16700059, ISI:000079102800040}.

Accessing the time-dependent trajectory of a micron-sphere, to high spatial and temporal resolution, is one of the basic principles behind microrheology techniques~\cite{ISI:A1995QG47200053, ISI:000274107900018}. Microrheology is a branch of \textit{rheology} (the study of flow of matter), but it works at micron length scales and with micro-litre sample volumes. Therefore, microrheology techniques are revealed to be very useful tools for all those rheological studies where rare or precious materials are employed, e.g.~in biophysical studies~\cite{ISI:000274013800006, ISI:000253676200021, ISI:000260776300073, ISI:000328623600017, ISI:000329157000012}. Moreover, microrheology measurements can be performed \textit{in situ} in an environment that cannot be reached by a bulk rheology experiment, for instance inside a living cell~\cite{ISI:000294028600004}.

However, while microrheology with OT of complex \textit{fluids} has been presented~\cite{ISI:000275053800036,ISI:000291926500023, T12} and validated~\cite{ISI:000322416700024} against conventional bulk rheology methods, when it is considered for rheological studies of soft \textit{solids} (e.g. gel-like materials or living cells), there exist some issues related to the viscoelastic nature of the compound system, made up of the optical trap and the complex \emph{solid}, that preclude the determination of the viscoelastic properties of such materials, as explained hereafter.

\section{Discussion}

\begin{figure}[t]
  \centering
  \includegraphics[width=10cm]{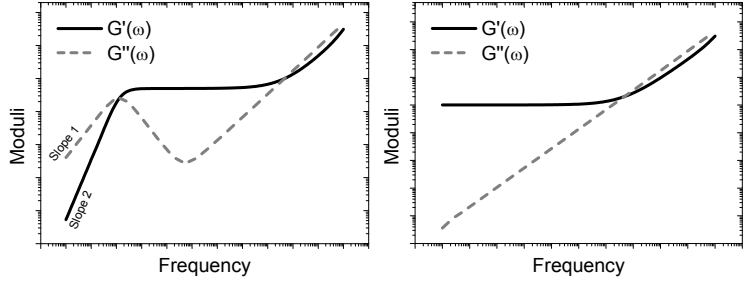}
\caption{Schematic representations of the frequency dependent moduli for a generic viscoelastic fluid (left) and a generic complex solid (right). Both the graphs have double logarithmic scales.}
\vspace{-1.5em}
\label{Fig1}
\end{figure}

As already introduced by Tassieri \textit{et al.}~\cite{T12} for the case of a stationary trap, the statistical mechanics analysis of the thermal fluctuations of an optically trapped micron-sized spherical particle, suspended in a \textit{fluid} at thermal equilibrium, has the potential of revealing both (i) the trap stiffness and (ii) the frequency-dependent viscoelastic properties of the suspending \emph{fluid}. The latter can be expressed in terms of the material's complex shear modulus $G^*(\omega)$; which is a complex number ($G^*(\omega)=G'(\omega)+iG''(\omega)$) whose real and imaginary parts provide information on the elastic and viscous nature of materials~\cite{Ferry:1980jo} (e.g. see Fig.~\ref{Fig1}), respectively.

The trap stiffness $\kappa$ can be determined by appealing to the Principle of Equipartition of Energy:
\begin{equation}
\label{Equipartition}
	\frac{3}{2}k_{B}T=\frac{1}{2}\kappa \langle r^2 \rangle
\end{equation}
where $k_{B}$ is the Boltzmann's constant, $T$ is the absolute temperature and $\langle r^2 \rangle$ is the time-independent variance of the particle's position $\vec{r}$ from the trap center, the origin of $\vec{r}$.
Despite the great variety of methods for determining the trap stiffness (e.g., using the power spectrum or the drag force~\cite{ISI:000342384600026}), Eq.~(\ref{Equipartition}) provides the only such measurement that is independent of the viscoelastic properties of the \emph{fluid} under investigation and is thus essential for proper calibration. This is because, whatever the elasticity of the unknown \emph{fluid}, its contribution to the time-independent constraining force must vanish at long times (because at rest the fluid's elastic shear modulus goes to zero as the time goes to infinity). This is equivalent to say that at low frequencies the elastic component of the viscoelastic fluid vanishes towards zero, as schematically shown in Fig.~\ref{Fig1} (left).~Thus the trap stiffness is easily determined by means of Eq.~(\ref{Equipartition}) applied to a \textit{sufficiently long measurement}~\cite{T12}.

Once calibrated, OT can be used to evaluate $G^*(\omega)$ by solving a generalised Langevin equation in terms of either the normalised mean square displacement $\Pi(\tau)=\langle\Delta r^2(\tau)\rangle / 2\left\langle r^2\right\rangle$~\cite{ISI:000275053800036} or the particle normalised position autocorrelation function $A(\tau)=\left\langle \vec{r}(t)\vec{r}(t+\tau)\right\rangle / \left\langle r^2\right\rangle$~\cite{ISI:000291926500023}:
\begin{equation}
\label{G*OT}
	G^*(\omega)\frac{6\pi a}{\kappa}=\frac{\hat{A}(\omega)}{\hat{\Pi}(\omega)}
\end{equation}
where $\tau$ is the lag-time, $a$ is the bead radius, $\hat{A}(\omega)$ and $\hat{\Pi}(\omega)$ are the Fourier transforms of $A(\tau)$ and $\Pi(\tau)$, respectively. Notice that, Eq.~(\ref{G*OT}) is valid as long as the particle inertia is negligible.

When a particle is embedded into a gel, Eq.~(\ref{Equipartition}) becomes underdetermined with two unknowns, $\kappa$ and $G'_0$:
\begin{equation}
\label{Equi2}
	\frac{3}{2}k_{B}T=\frac{1}{2}\kappa\left(1+G'_0\frac{6\pi a}{\kappa}\right) \langle r^2 \rangle
\end{equation}

\noindent This because the elastic component of the viscoelastic \emph{solid} does not vanish at long times (i.e. low frequencies), but actually it approaches a frequency-independed plateau value ($G'_0$), as schematically shown in Fig.~\ref{Fig1} (right). Thus the inappropriateness of OT for microrheology of gels, as it is not possible (\emph{yet}) to discriminate between the two elastic contributions.

In the case of living cells, things get more complicated, because their mechanical properties change with time~\cite{ISI:000243535400040}, especially at low frequencies (i.e. long times). Nevertheless, microrheology of living organisms can still be performed if the right assumptions on the characteristic times involved during the measurement are made; i.e., on the Deborah Number ($De=time~of~relaxation/time~of~observation$)~\cite{reiner1964deborah}. Indeed, for each organism, one could assume the existence of a characteristic time ($\tau_{org}$) such that, for measurements having duration $T_m$ shorter than $\tau_{org}$ (i.e., $De\gtrsim1$), the living system can be seen as a complex material (either \emph{fluid} or \emph{solid}) with \emph{`time-invariant'} viscoelastic properties; whereas, for observations lasting longer than $\tau_{org}$ (i.e., $De\lesssim1$), the living organism has time to self-reorganise and thus to change its mechanical properties.
Therefore, notwithstanding that for $De\gtrsim1$ microrheology measurements of a living system are possible; in the case of OT, the following two considerations exclude them from being considered for such purpose. In particular, (i) if the living organism is assumed to behave as a complex \emph{fluid}, then OT would require a \textit{sufficiently long measurement} for a proper calibration of the trap stiffness, but this would result in $De\lesssim1$ with the lost of the pseudo-equilibrium assumption because of the self-reorganising process of the organism; (ii) if the latter is assumed to behave as a complex \emph{solid}, then the considerations made above for the gels would hold and OT would not be able to provide information on the absolute magnitude of $G^*(\omega)$, but only information on the frequency behaviour of its real and imaginary parts that anyway could still provide some useful insights on the system under investigation~\cite{ISI:000334027300022}.

In conclusion, based on simple rheological considerations, it has been shown that microrheology with optical tweezers of gel-like materials `\textit{is not an option}'!

\bibliographystyle{unsrt}    
\bibliography{OSA}

\begin{thebibliography}{10}

\bibitem{Ashkin:1970}
A.~Ashkin.
\newblock Acceleration and trapping of particles by radiation pressure.
\newblock {\em Phys. Rev. Lett.}, 24(4):156--159, 1970.

\bibitem{Bowman:2013}
Richard~W Bowman and Miles~J Padgett.
\newblock Optical trapping and binding.
\newblock {\em Reports on Progress in Physics}, 76(2):026401, 2013.

\bibitem{ISI:A1989U008400059}
SM~Block, DF~Blair, and HC~Berg.
\newblock {Compliance of bacterial flagella measured with optical tweezers}.
\newblock {\em {Nature}}, {338}({6215}):{514--518}, {APR 6} {1989}.

\bibitem{ISI:A1994NA03000055}
JT~Finer, RM~Simmons, and JA~Spudich.
\newblock {Single myosin molecule mechanics - piconewton forces and nanometer
  steps}.
\newblock {\em {Nature}}, {368}({6467}):{113--119}, {MAR 10} {1994}.

\bibitem{ISI:000259700300007}
Young-Zoon Yoon, Jurij Kotar, Gilwon Yoon, and Pietro Cicuta.
\newblock {The nonlinear mechanical response of the red blood cell}.
\newblock {\em {Physical biology}}, {5}({3}), {SEP} {2008}.

\bibitem{ISI:A1987G470500040}
A.~Ashkin and J.~M. Dziedzic.
\newblock {Optical trapping and manipulation of viruses and bacteria}.
\newblock {\em {Science}}, {235}({4795}):{1517--1520}, {MAR 20} {1987}.

\bibitem{ISI:A1997WZ16700059}
L~Tskhovrebova, J~Trinick, JA~Sleep, and RM~Simmons.
\newblock {Elasticity and unfolding of single molecules of the giant muscle
  protein titin}.
\newblock {\em {Nature}}, {387}({6630}):{308--312}, {MAY 15} {1997}.

\bibitem{ISI:000079102800040}
AD~Mehta, M~Rief, JA~Spudich, DA~Smith, and RM~Simmons.
\newblock {Single-molecule biomechanics with optical methods}.
\newblock {\em {Science}}, {283}({5408}):{1689--1695}, {MAR 12} {1999}.

\bibitem{ISI:A1995QG47200053}
TG~Mason and DA~Weitz.
\newblock {Optical measurements of frequency-dependent linear viscoelastic
  moduli of complex fluids}.
\newblock {\em {Physical Review Letters}}, {74}({7}):{1250--1253}, {FEB 13}
  {1995}.

\bibitem{ISI:000274107900018}
Todd~M. Squires and Thomas~G. Mason.
\newblock {Fluid Mechanics of Microrheology}.
\newblock {\em {Annual review of fluid mechanics}}, {42}:{413--438}, {2010}.

\bibitem{ISI:000274013800006}
M~Tassieri, TA~Waigh, John Trinick, Amalia Aggeli, and RML Evans.
\newblock {Analysis of the linear viscoelasticity of polyelectrolytes by
  magnetic microrheometry-Pulsed creep experiments and the one particle
  response}.
\newblock {\em Journal of rheology}, {54}({1}):{117--131}, {JAN-FEB} {2010}.

\bibitem{ISI:000253676200021}
M.~Tassieri, RML Evans, L~Barbu-Tudoran, J~Trinick, and TA~Waigh.
\newblock {The self-assembly, elasticity, and dynamics of cardiac thin
  filaments}.
\newblock {\em {Biophysical journal}}, {94}({6}):{2170--2178}, {MAR 15} {2008}.

\bibitem{ISI:000260776300073}
Manlio Tassieri, RML Evans, Lucian Barbu-Tudoran, G~Nasir Khaname, John
  Trinick, and Tom~A Waigh.
\newblock {Dynamics of Semiflexible Polymer Solutions in the Highly Entangled
  Regime}.
\newblock {\em {Physical Review Letters}}, {101}({19}), {NOV 7} {2008}.

\bibitem{ISI:000328623600017}
Fiona Watts, Lay~Ean Tan, Clive~G. Wilson, John~M. Girkin, Manlio Tassieri, and
  Amanda~J. Wright.
\newblock {Investigating the micro-rheology of the vitreous humor using an
  optically trapped local probe}.
\newblock {\em {Journal of Optics}}, {16}({1}), {JAN} {2014}.

\bibitem{ISI:000329157000012}
Emma~J. Robertson, Grace Najjuka, Melissa~A. Rolfes, Andrew Akampurira, Neena
  Jain, Janani Anantharanjit, Maximilian von Hohenberg, Manlio Tassieri, Allan
  Carlsson, David~B. Meya, Thomas~S. Harrison, Bettina~C. Fries, David~R.
  Boulware, and Tihana Bicanic.
\newblock {Cryptococcus neoformans Ex Vivo Capsule Size Is Associated With
  Intracranial Pressure and Host Immune Response in HIV-associated Cryptococcal
  Meningitis}.
\newblock {\em {Journal of Infectious Diseases}}, {209}({1}):{74--82}, {JAN 1}
  {2014}.

\bibitem{ISI:000294028600004}
Philip Kollmannsberger and Ben Fabry.
\newblock {Linear and Nonlinear Rheology of Living Cells}.
\newblock {\em {Annual review of materials research}}, {41}:{75--97}, {2011}.

\bibitem{ISI:000275053800036}
Manlio Tassieri, Graham~M Gibson, RML Evans, Alison~M Yao, Rebecca Warren,
  Miles~J Padgett, and Jonathan~M Cooper.
\newblock {Measuring storage and loss moduli using optical tweezers: Broadband
  microrheology}.
\newblock {\em {Physical Review E}}, {81}({2, Part 2}), {FEB} {2010}.

\bibitem{ISI:000291926500023}
Daryl Preece, Rebecca Warren, RML Evans, Graham~M Gibson, Miles~J Padgett,
  Jonathan~M Cooper, and Manlio Tassieri.
\newblock {Optical tweezers: wideband microrheology}.
\newblock {\em Journal of optics}, {13}({4, SI}), {APR} {2011}.

\bibitem{T12}
Manlio Tassieri, RML Evans, Rebecca~L Warren, Nicholas~J Bailey, and Jonathan~M
  Cooper.
\newblock Microrheology with optical tweezers: data analysis.
\newblock {\em New Journal of Physics}, 14(11):115032, 2012.

\bibitem{ISI:000322416700024}
Angelo Pommella, Valentina Preziosi, Sergio Caserta, Jonathan~M. Cooper,
  Stefano Guido, and Manlio Tassieri.
\newblock {Using Optical Tweezers for the Characterization of Polyelectrolyte
  Solutions with Very Low Viscoelasticity}.
\newblock {\em {Langmuir}}, {29}({29}):{9224--9230}, {JUL 23} {2013}.

\bibitem{Ferry:1980jo}
John~D. Ferry.
\newblock {\em Viscoelastic properties of polymers}.
\newblock Wiley, 3d ed edition, 1980.

\bibitem{ISI:000342384600026}
Yonggun Jun, Suvranta~K. Tripathy, Babu R.~J. Narayanareddy, Michelle~K.
  Mattson-Hoss, and Steven~P. Gross.
\newblock {Calibration of Optical Tweezers for In Vivo Force Measurements: How
  do Different Approaches Compare?}
\newblock {\em {Biophysical Journal}}, {107}({6}):{1474--1484}, {SEP 16}
  {2014}.

\bibitem{ISI:000243535400040}
Daisuke Mizuno, Catherine Tardin, C.~F. Schmidt, and F.~C. MacKintosh.
\newblock {Nonequilibrium mechanics of active cytoskeletal networks}.
\newblock {\em {Science}}, {315}({5810}):{370--373}, {JAN 19} {2007}.

\bibitem{reiner1964deborah}
M~Reiner.
\newblock The deborah number.
\newblock {\em Physics today}, 17(1):62, 1964.

\bibitem{ISI:000334027300022}
Rebecca~L. Warren, Manlio Tassieri, Xiang Li, Andrew Glidle, David~J. Paterson,
  Allan Carlsson, and Jonathan~M. Cooper.
\newblock {Rheology at the micro-scale: new tools for bio-analysis}.
\newblock In {\em {Optical Methods for Inspection, Characterization, and
  Imaging of Biomaterials}}, volume {8792} of {\em {Proceedings of SPIE}}.
  {SPIE}, {2013}.

\end{thebibliography}

\end{document}